%% file: ms_astroph.tex
\documentclass[rnote]{aa}

\usepackage{txfonts}
\usepackage{natbib}
\usepackage{graphicx}

\begin{document}

\title{Bianchi Type VII$_h$ Models and the \emph{WMAP} 3-year Data}

\author{T.\ R. Jaffe\inst{1} \and 
A.\ J.\ Banday\inst{1} \and 
H.\ K.\ Eriksen\inst{2} \and 
K.\ M.\ G\'orski\inst{3,4}  \and 
F. K. Hansen\inst{2}
}

\institute{
Max-Planck-Institut f\"ur Astrophysik,
Karl-Schwarzschild-Str.\ 1, Postfach 1317, D-85741 Garching bei
M\"unchen, Germany; tjaffe@MPA-Garching.MPG.DE, 
banday@MPA-Garching.MPG.DE.  
\and
Institute of Theoretical Astrophysics, University of
Oslo, P.O.\ Box 1029 Blindern, N-0315 Oslo, Norway; 
h.k.k.eriksen@astro.uio.no, f.k.hansen@astro.uio.no. 
\and 
JPL, M/S 169/327, 4800 Oak Grove Drive, Pasadena CA
91109; Krzysztof.M.Gorski@jpl.nasa.gov
\and Warsaw University Observatory, Aleje Ujazdowskie 4, 00-478
Warszawa, Poland
}

\date{Received <date>  / Accepted <date> }

\abstract 
{A specific example of Bianchi Type VII$_h$ models, i.e. those including universal rotation
  (vorticity) and differential expansion (shear), has been shown in
  Jaffe et al. (2005) to correlate unexpectedly with the \emph{WMAP}
  first-year data.}
{We re-assess the signature of this model in the \emph{WMAP} 3-year
  data.}
{The cross-correlation methods are described in Jaffe et al. (2006a).
  We use the \emph{WMAP} 3-year data release, including maps for
  individual years, and perform additional comparisons to assess the
  influence of both noise and residual foregrounds and eliminate
  potential non-cosmological sources for the correlation.}
{We confirm that the signal is detected in both the combined 3-year data
and the individual yearly sky maps at a level consistent with our
original analysis. The significance of the correlation is not
  affected by either noise or foreground residuals.  }
{The results of our previous study are unchanged.}

\keywords{cosmic microwave background --- cosmology: observations }

\maketitle

\section{Introduction}

\label{sec:introduction}

In \citet{jaffe:2005,jaffe:2006a}, we reported on the unexpected
detection of a correlation between the cosmic microwave background 
(CMB) sky measured by the \emph{WMAP}
first-year data release and a Bianchi Type VII$_h$ template.  
This result is particularly provocative 
in that such a signal may provide an explanation for several
independently discovered anomalies: the low quadrupole amplitude
\citep{doc:2004} relative to the best-fit cosmological model, the
curious alignment of several large angular scale multipoles along the
so-called ``axis of evil'' \citep{land:2005a}, a significant power
asymmetry between two hemispheres on the sky
\citep{hansen:2004a,eriksen:2004a} and a large temperature decrement
towards a specific part of the southern sky as revealed by a wavelet
analysis \citep{cruz:2005,vielva:2004}.  The subtraction of the
Bianchi component leaves a statistically isotropic sky.

In this note, we examine the newly
released three-year data described in \cite{hinshaw:2006} and show
that the signal is detected both in the combined 3-year data and
in each individual yearly map, despite differences in the data processing
pipelines, calibration and corrections for foreground contamination.

\section{Methods}

The methods used are described in detail in \citet{jaffe:2006a}.  In
brief, we use a minimum-$\chi^2$ method to estimate the
cross-correlation amplitude between the \emph{WMAP} data and the
deterministic Bianchi template derived by \citet{barrow:1985} and
extended in \citet{jaffe:2006b} to include arbitrary dark energy
models.  

Full sky analysis is only possible using the highly processed 
internal linear combination (ILC) map, created by combining the 
data at different frequencies using weights
that minimize the variance while preserving the cosmic signal.  
Such a combination is expected to clean the data of galactic emission.
However, as pointed out by \citet{eriksen:2004b}, the
method does leave significant foreground residuals which particularly 
affect the galactic plane region.
\citet{hinshaw:2006} point out that this is a consequence of the
method minimising the variance by anti-correlating the
CMB and foregrounds.
In the 3-year release, an additional ``de-biasing'' is therefore
applied that attempts to correct for this.
With these data,
we use the efficient total convolver algorithm \citep{wandelt:2001} in
harmonic space to find the best orientation of the Bianchi template relative
to the data. We then use the sky maps from ten individual differencing
assemblies (DAs) spanning five frequencies (K, Ka, Q, V, W) to confirm this result
and assess the extent to which the full sky fit is affected by foreground
residuals.

For individual frequency bands (derived by co-adding the corresponding
DAs together) and combinations of bands, we fix the template at its
best-fit location and perform simultaneous fits in pixel space of the
Bianchi template and three templates to model the galactic foreground
contribution.  
For thermal dust emission, we use either  
the 100$\mu$m sky map derived by Schlegel et al. (1998, hereafter SFD)
or a template at an effective frequency of 94 GHz (the \emph{WMAP}
W-band) derived from SFD using model 8 of Finkbeiner et al. 
(1999, hereafter FDS) that includes an additional correction for the
dust temperature.
Free-free emission is traced by maps of H$_\alpha$. We consider two
templates that differ in the way that they combine the available
H$_\alpha$ survey data -- Dickinson et al. (2003, hereafter DDD) or
\citet{Finkbeiner:2003}. 
For Galactic synchrotron emission we again consider two
alternative templates: the 408 MHz survey of \cite{haslam:1982},
and, following the \emph{WMAP} analysis, the K-Ka difference map. 
The advantage of the latter is that it will more realistically 
account for the synchtron emission at the frequencies of interest,
capturing some variation in the synchrotron spectral 
behaviour that a simple extrapolation of the 408 MHz data cannot.
Moreover, it will also trace {\em any} additional foreground component 
at low frequencies that is not represented by the dust or H$_\alpha$
templates.
This includes at least some fraction of the co-called ``anomalous dust'' emission --
due to its correlation with the dust template -- although
the physical origin of this component still remains unclear (see
\citealt{davies:2006} for a relevant discussion).
The fits are confined 
to the region outside the conservative Kp0 mask that eliminates bright
emission from the Galactic plane and several hundred point sources. 
This is necessary since the former exhibits complex spatially varying
spectral behaviour that the templates are not sufficiently
accurate to quantify. 
We estimate the effect of high-latitude foreground residuals not well traced by 
these templates by fitting the Bianchi models to the foreground
maps derived in \citet{hinshaw:2006} using a maximum-entropy (MEM)
method. These include pixel-by-pixel variations in the foreground
emission, but their noise properties are usual considered too complex
to include in a cosmological analysis. However, this is not an issue
for this work.

In order to characterize the effect of noise and systematic artifacts
that are not well described by the \emph{WMAP} noise model, we utilize
year-to-year difference maps. 

Finally, we use the LILC simulation pipeline of \citet{eriksen:2004b} to
characterize the null hypothesis, namely the distribution of best-fit
chance alignments arising between the Bianchi template and simulated
skies containing no such contribution.  A significance measure for a given template is
then determined from the number of chance alignments with higher amplitude
than that measured using the data.

For computational efficiency, all maps are smoothed to a 5$\degr$.5 FWHM beam and downgraded to
HEALPix\footnote{\tt{http://healpix.jpl.nasa.gov/}} N$_{\rm side}=32$
resolution.  All pixel-space fits include simultaneous fits of
the monopole and dipole components, since these may contain additional
non-cosmological contributions.

\section{Results}

\begin{figure}[tl]
\resizebox{\hsize}{!}{\includegraphics[width=88mm]{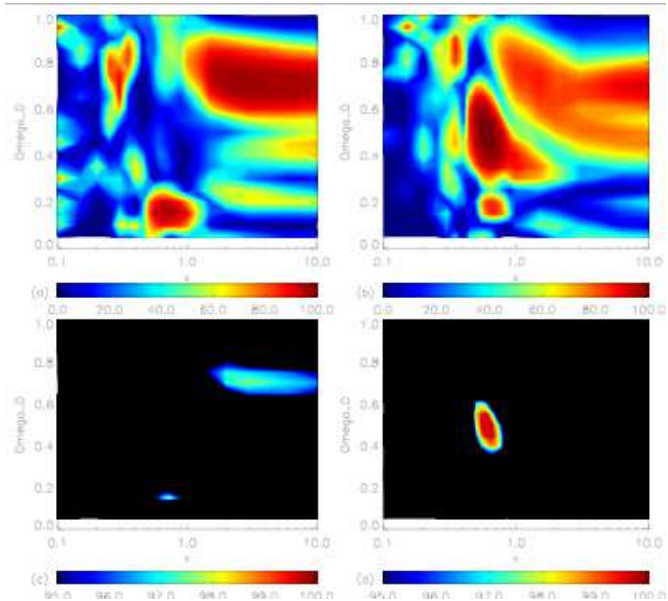}}
\caption{Significance contours as percentage of LILC simulations whose
best-fit chance alignment amplitude is lower.  The left (right) column
shows the left-handed (right-handed) models.  Over plotted contours are at
$99.3$ and $99.5\%$.  Two color scales are used to show the global
structure (top) as well as that near the peaks
(bottom).\label{fig:contours}
}
\end{figure}

The results of the model-space search using the ILC map over the full
sky are presented in Fig.~\ref{fig:contours}.  As in
\citet{jaffe:2006a}, there is a region of high significance in the
parameter space for right-handed models around
$(x,\Omega_0)=(0.6,0.5)$ seen in the right-hand column of
Fig.~\ref{fig:contours}.  In the previous work, there was also a
possible detection of a left-handed model at
$(x,\Omega_0)=(0.6,0.15)$, which we examined in detail as well and
concluded was due to foreground contamination in the galactic plane
region.  The left-hand column of Fig.~\ref{fig:contours} shows that
this model is not a significant fit to the 3-year data.  This change
is likely due to the differences introduced by the de-biasing applied
to the ILC in the 3-year release.  This analysis repeats the search for
the best-fit orientation of the template relative to the data and
finds the same position as the previous best-fit to within one bin of
2.8$\degr$ size at this resolution.  Using the full-sky ILC maps, the
best-fit model amplitude drops a few percent from $4.33\times
10^{-10}$ to $4.19\times 10^{-10}$, which corresponds to a change in
significance from $99.8\%$ to $99.5\%$.

%

\begin{figure}
\resizebox{88mm}{!}{\includegraphics{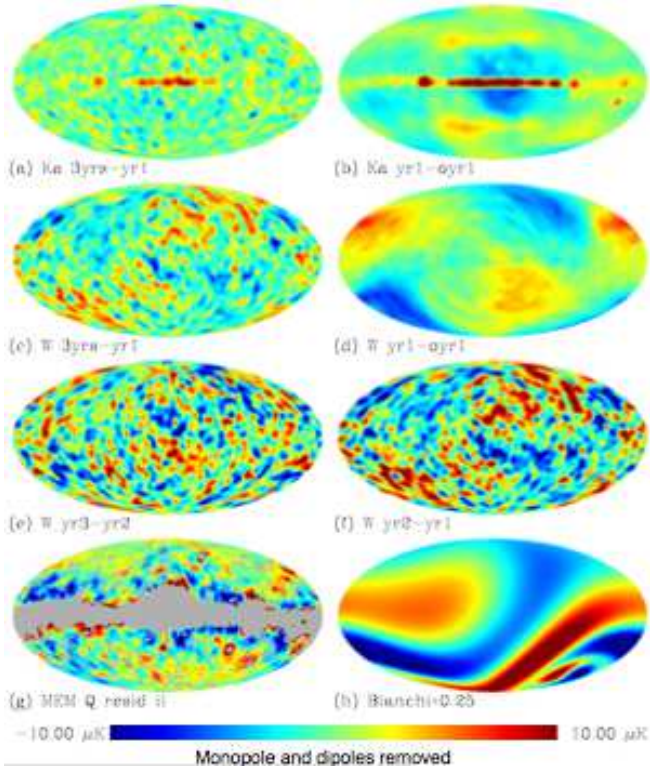}}
\caption{ Difference maps showing noise and calibration effects.  
The monopole and dipole terms are removed. See text for details.
\label{fig:diff_maps}
}
\end{figure}

Fig.~\ref{fig:diff_maps} presents several difference maps that
indicate the amplitude and morphology of noise, calibration, and map-making effects
which might influence the analysis.    
Fig.~\ref{fig:diff_maps}~(a) shows the Ka-band difference
between the 3-year and 1-year map, where the time variability in a few
point sources (shown here highly smoothed) is the dominant component.
Fig.~\ref{fig:diff_maps}~(b) depicts the 
differences in the 1$^{\rm st}$ year Ka-band sky maps
arising from changes in the map-making and calibration methods
between the current and original data releases. The observed structure
arises from changes in the gain calibration solutions, 
and is particularly notable in the smoothed differences around point-sources along the
galactic plane.
See \citet{jarosik:2006} for a discussion of these differences.
Figs.~\ref{fig:diff_maps}~(c) and (d) are the same for the W-band,
although the point sources are no longer a significant contaminant.
Figs.~\ref{fig:diff_maps}~(e) and (f) show two examples of the noise
differences in the W-band between pairs of individual years.
Fig.~\ref{fig:diff_maps}~(g) gives the residuals from fitting the
foreground templates to the MEM Q-band foregrounds (see below).
Lastly, Fig.~\ref{fig:diff_maps}~(h) shows the Bianchi template at one quarter
of its best-fit amplitude for comparison.  The Bianchi anisotropy
amplitude is clearly much higher than any of these effects.

\begin{figure}
\resizebox{88mm}{!}{\includegraphics{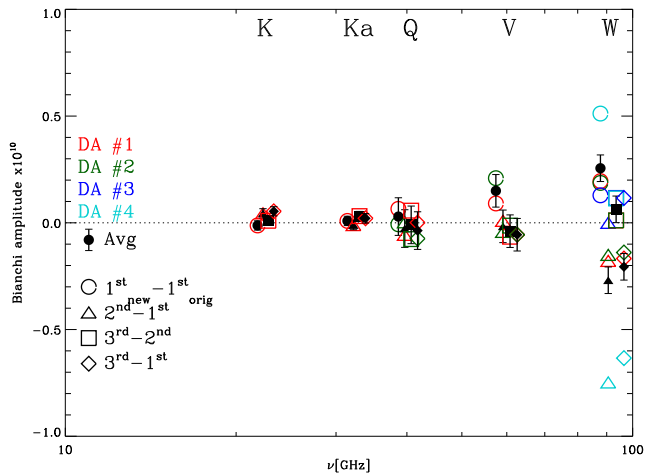}}
\caption{ Fits to year-to-year difference maps by assembly.  The
different symbols show different pairs of years, while the color
indicates the differencing assembly.  The filled black symbols are the
averages over the different assemblies.
\label{fig:bspectra_das}}
\end{figure}

\begin{figure}
\resizebox{88mm}{!}{\includegraphics{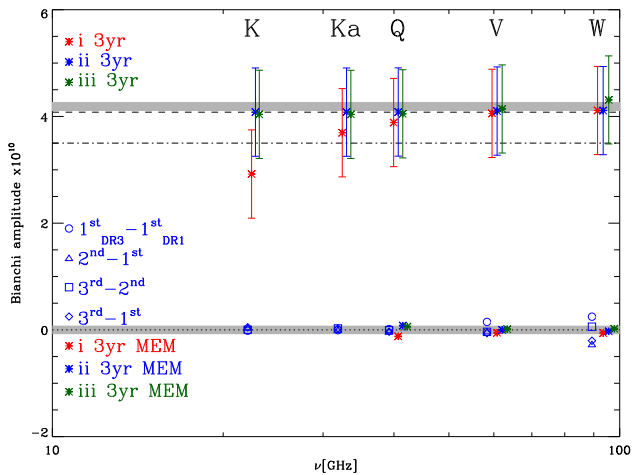}}
\caption{ \emph{WMAP} 3-year fit amplitudes for the three different sets of templates labeled columns {\it i}, {\it ii}, and {\it iii} in Table \ref{tab:fits_summary}.  Also plotted are fit results from difference maps between years and from the MEM foreground maps.  These indicate the degree to which noise and foreground residuals may be contaminating the fits, shown as the grey band.  This band is also shown around the best-fit ILC amplitude for comparison.  The dashed and dot-dashed lines represent the approximate amplitudes corresponding to $99\%$ and $95\%$ significance, respectively.
\label{fig:bspectra} }
\end{figure}

In Fig.~\ref{fig:bspectra_das}, we plot the fits to year-to-year
difference maps for each DA.  The
scatter clearly increases at high frequency, as expected due to the
increased noise.  W4 in particular shows significantly higher scatter
than the other W-band assemblies, which is unsurprising considering
that it has the highest level of $1/f$ noise (see
\citealt{jarosik:2003}).  We have also performed fits to maps
constructed from the MEM foreground solutions to test for the effect
of residuals not traced by the foreground templates.  We find that the maximum
Bianchi amplitude as fitted to these maps is less than 3\% of the ILC
fit amplitude, highest for the Q-band fits as expected for foregrounds
that increasingly dominate at lower frequencies.  The fits are plotted
in Fig.~\ref{fig:bspectra}, and the grey bands reflect the
uncertainty in the fit results due to the noise and foreground
residuals.  The error bars are the statistical uncertainties almost
entirely due to the isotropic CMB contribution.

Table~\ref{tab:fits_summary} shows that the results of the Bianchi
fitting to the \emph{WMAP} data are very stable 
and largely robust to foreground correction method and template selection.
Although the fits to individual frequency bands are slightly lower than
the ILC fit amplitude, they remain near or above the $99\%$
significance level. The most consistent results are derived
when the K-Ka map is used as a synchrotron tracer, 
presumably because it can also account for whatever
low-frequency foreground emission is not traced by the dust and
free-free templates.
By comparison, using the template combination of FDS dust,
Finkbeiner H$_\alpha$, and Haslam for synchrotron (as in our original
analysis), there remains some low-frequency residual which
anti-correlates with the CMB and therefore lowers the fit amplitudes.
Using the DDD instead of the Finkbeiner H$_\alpha$
template and the SFD instead of the FDS thermal dust template again
yields very consistent results.  There is a small difference at W-band
where the SFD template perhaps does not account for the thermal dust
emission as well as the FDS template does.  

The difference maps, which have no cosmic signal, give fit amplitudes
of only 3\% or less of the signal, consistent with the estimate given
by the MEM fits. The 3-year fit values are
also plotted in Fig.~\ref{fig:bspectra} with fits to noise and
foreground maps additionally shown for comparison. The difference between the
3-year fits and the individual years is too small to see on this
plot. The grey band around the fit value obtained with the ILC shows
that, though they are slightly lower, the cut-sky fits do not vary in
excess of what is expected due to such small foreground residuals and
noise.  Even with these effects, the fits are consistently at the 99\%
significance level compared with simulations (shown by the dashed
line).

%
\input{table}

\section{Conclusions}

In \citet{jaffe:2005,jaffe:2006a}, we showed that removing a Bianchi component from the 
\emph{WMAP} initial data release can account for several
large angular scale anomalies and yield a corrected sky
that is statistically isotropic. 
Although we do not repeat all of these analyses here,
we simply note that the change in the fit
amplitude is too small to alter the previous conclusions.
We have additionally shown in this work that the correlation cannot be
attributed either to instrument noise, foreground residuals, or any
known systematic corrected in the three-year \emph{WMAP} data release.

In \citet{jaffe:2006b}, we
found that the cosmological parameters required to reproduce the
Bianchi morphology are not compatible with the ``cosmic concordance''
of other datasets, including the small-scale CMB structure itself,
ruling out these models as a realistic physical explanation for the
anomalies. Nevertheless, a recent analysis with a statistically rigorous method
in \citet{bridges:2006} has
demonstrated, in agreement with our results,  that the data do
appear to require the Bianchi template.  They give the ``Bayesian evidence''
in favor of a cosmology including vorticity and shear compared to one
without as roughly $\Delta ln E\sim2$, which is considered
``significant''.  

We conclude, therefore, that in the absence of an unknown systematic
effect which could explain both the anomalies and the correlation, the
\emph{WMAP} data require an addition to the standard cosmological
model that resembles the Bianchi morphology.

\begin{acknowledgements}
We acknowledge use of the HEALPix software \citep{healpix} and of the
Legacy Archive for Microwave Background Data Analysis (LAMBDA).
Support for LAMBDA is provided by the NASA Office of Space Science.
\end{acknowledgements}

\bibliographystyle{aa}

\bibliography{jabref}


\end{document}

%% file: table.tex
%
%
\begin{table*}

\begin{tabular}{cccccccccccccccccccc}
\hline\hline
 Map & \multicolumn{12}{c}{Amplitudes $(\sigma/H)_0$ $\times 10^{10}$} & \multicolumn{3}{c}{$P(|\alpha_{\rm sim}| < |\alpha_{\rm obs}| ) $}\\

  & \multicolumn{3}{c}{$1^{\rm st}$-{\it year}} & \multicolumn{3}{c}{$2^{\rm nd}$-{\it year}} & \multicolumn{3}{c}{$3^{\rm rd}$-{\it year} } & \multicolumn{3}{c}{3-year} & \multicolumn{3}{c}{3-year \%}  \\
\hline
WILC & \multicolumn{9}{c}{} & \multicolumn{3}{c}{$ 4.19\pm 0.82$} & \multicolumn{3}{c}{$99.50$} \\

\hline
 & {\it i} & {\it ii} & {\it iii} & {\it i} & {\it ii} & {\it iii} & {\it i} & {\it ii} & {\it iii} & {\it i} & {\it ii} & {\it iii} & {\it i} & {\it ii} & {\it iii} \\ 
\hline

K &  2.92 &  4.08 &  4.03 &  2.96 &  4.12 &  4.08 &  2.95 &  4.10 &  4.08 &  2.92 &  4.08 &  4.04 &  41.6 &  98.9 &  98.8\\
KA &  3.68 &  4.08 &  4.03 &  3.70 &  4.09 &  4.05 &  3.70 &  4.08 &  4.04 &  3.69 &  4.08 &  4.04 &  93.9 &  99.0 &  98.8\\
Q &  3.87 &  4.07 &  4.03 &  3.88 &  4.07 &  4.04 &  3.91 &  4.10 &  4.07 &  3.89 &  4.08 &  4.05 &  97.3 &  99.1 &  98.9\\
V &  4.09 &  4.13 &  4.17 &  4.06 &  4.10 &  4.14 &  4.03 &  4.07 &  4.11 &  4.06 &  4.10 &  4.14 &  98.9 &  99.2 &  99.3\\
W &  4.15 &  4.15 &  4.35 &  4.06 &  4.05 &  4.26 &  4.13 &  4.13 &  4.33 &  4.11 &  4.11 &  4.31 &  99.1 &  99.1 &  99.8\\
QVW &  3.95 &  4.10 &  4.10 &  3.94 &  4.08 &  4.09 &  3.95 &  4.09 &  4.10 &  3.95 &  4.09 &  4.09 &  98.0 &  99.1 &  99.1\\
VW &  4.10 &  4.14 &  4.20 &  4.06 &  4.09 &  4.17 &  4.05 &  4.08 &  4.15 &  4.07 &  4.10 &  4.17 &  98.9 &  99.2 &  99.3\\
Q-V & -0.06 &  0.08 &  0.05 & -0.07 &  0.07 &  0.04 & -0.05 &  0.10 &  0.07 & -0.06 &  0.08 &  0.05 &  - &  - &  -\\
V-W & -0.13 & -0.09 & -0.16 &  0.11 &  0.15 &  0.08 &  0.03 &  0.06 & -0.01 &  0.00 &  0.04 & -0.03 &   - &  - &  - \\
Q-W & -0.20 & -0.01 & -0.11 &  0.04 &  0.22 &  0.12 & -0.02 &  0.16 &  0.06 & -0.06 &  0.12 &  0.02 &  - & - &  - \\
\hline
\end{tabular}
 \caption{Amplitudes
(i.e. the shear value $(\sigma/H)_0\times 10^{10}$) of the
best-fit model derived from various combinations of data and various
methods as described in the text of \citet{jaffe:2006a}.
Uncertainties are shown for the ILC only, and they are roughly the
same for all, since the noise is negligible for this low-resolution
analysis and the structure outside the Galactic cut.  For the WILC
maps, the total convolver method is used on the full sky.  For the
individual bands, the Kp0 mask was imposed and foreground templates
fit simultaneously for the remaining maps: in columns labeled {\it
(i)}, we used the FDS dust, Finkbeiner H$_\alpha$, and Haslam for
synchrotron; in columns labeled {\it (ii)}, we used \emph{WMAP} K-Ka
for synchrotron; in columns labeled {\it (iii)}, we use alternate
templates SFD for dust and DDD for H$_\alpha$, with again
\emph{WMAP} K-Ka for synchrotron.  The last columns show the
significance measure for the 3-year results, i.e. the percentage of
simulations with lower amplitude. 
The significances of the difference map fits 
are not quoted since they would reflect
the fact that there are more residuals in the data 
than in idealised simulations
where the fitted templates are themselves used to generate the
foreground contribution. 
\label{tab:fits_summary}}
\end{table*}